\documentclass[a4paper,fleqn]{cas-dc}
\usepackage[authoryear]{natbib}
\let\cite\citep
\usepackage{graphicx}
\usepackage{amsmath}
\usepackage{amssymb}
\usepackage{booktabs}
\usepackage{array}
\usepackage{placeins}
\hypersetup{hidelinks}
\usepackage{xcolor}
\usepackage[normalem]{ulem}
\usepackage{tikz}
\usetikzlibrary{arrows.meta, positioning, calc, fit, backgrounds}


\begin{document}
\let\WriteBookmarks\relax
\shorttitle{Contrastive Astromer}
\shortauthors{Justin Minggan Wei, Pavlos Protopapas, Guillermo Cabrera-Vives}
\title[mode=title]{Contrastive Astromer: Two-Stage Contrastive Representation Learning for Astronomical Light Curve Classification}
\author[1]{Justin Minggan Wei}[orcid=0009-0009-4239-7070]
\cormark[1]
\ead{mwei@fas.harvard.edu}
\author[1]{Pavlos Protopapas}[orcid=0000-0002-8178-8463]
\author[2]{Guillermo Cabrera-Vives}[orcid=0000-0002-2720-7218]

\affiliation[1]{organization={School of Engineering and Applied Sciences, Harvard University},
                city={Cambridge},
                state={MA},
                postcode={02138},
                country={USA}}
\affiliation[2]{organization={Department of Computer Science, Universidad de Concepci\'on},
                addressline={Edmundo Larenas 219},
                city={Concepci\'on},
                country={Chile}}
\cortext[1]{Corresponding author}

\begin{abstract}
Large astronomical surveys collect abundant light curves, but reliable class labels remain scarce. ASTROMER 2 learns from unlabeled light curves by reconstructing masked observations, yet this objective does not explicitly encourage class separation in a compact representation of each light curve. We introduce Contrastive Astromer, a two-stage approach that makes these representations more useful for classification without changing the ASTROMER 2 encoder architecture. First, unsupervised contrastive learning on single-band MACHO light curves brings together representations of augmented views of the same object. Supervised contrastive learning, supported by an auxiliary classification loss, then uses a separate labeled Alcock sample to strengthen class discrimination. We evaluate the resulting representations by freezing the encoder and training a separate classifier. On a held-out set of 600 objects, this classifier achieves a macro-F1 of 0.7943 with 500 labeled examples per class, exceeding our reproduction of the ASTROMER 2 finetune-and-classify pipeline by 5.10 percentage points. With downstream classifier training budgets of 20 and 100 examples per class, macro-F1 reaches 0.7078 and 0.7704, respectively, also surpassing this baseline. Embedding-space analyses further show that supervised adaptation improves local class consistency and global class separation. These results show that contrastive adaptation makes pretrained light-curve representations more effective for classification and more informative about the structure of variability classes.
\end{abstract}


\maketitle

\section{Introduction}
\label{sec:introduction}

Light curves provide a record of stellar variability, but turning those records into reliable classifications remains challenging. Their periods, amplitudes, and shapes carry information about variability class and the underlying stellar processes~\cite{eyer2008variable,debosscher2007automated}. Wide-field surveys such as ZTF and ATLAS provide large collections of these observations~\cite{bellm2019ztf,tonry2018atlas}. Yet light curves are irregularly sampled, noisy, and often incomplete, while reliable class labels remain scarce. These challenges motivate models that learn from large unlabeled archives and produce representations that can be reused for classification~\cite{donoso2026astromer2,bommasani2021opportunities}.

Masked reconstruction enables light-curve encoders to learn from observations without requiring class labels. ASTROMER and ASTROMER~2 use this approach by hiding some observed magnitudes during training and predicting them from the remaining temporal context~\cite{donoso2023astromer,donoso2026astromer2}. The encoder produces a sequence of feature vectors, one for each input observation. A simple way to obtain a fixed-length summary is to average these vectors over the visible observations within an input window. We refer to this summary as the \emph{pooled representation}. This vector provides a convenient input to a classifier, but the reconstruction objective does not explicitly require it to distinguish between variability classes.

Contrastive learning offers a way to encourage useful relationships between representations during training. It brings representations of selected positive pairs closer together while separating them from negative pairs~\cite{oord2018representation,chen2020simple}. For example, two independently perturbed versions of the same input can form a positive pair, allowing the model to learn which features should remain consistent across those perturbations. When labels are available, different objects from the same class can also form positive pairs~\cite{khosla2020supervised}. We therefore investigate whether contrastive adaptation can make the pooled ASTROMER~2 representation more useful for variable-star classification while preserving the encoder architecture and single-band input.

We introduce \emph{Contrastive Astromer}, a two-stage approach that combines learning from unlabeled light curves with supervised adaptation. Starting from the pretrained ASTROMER~2 encoder, Stage~1 applies unsupervised contrastive learning to MACHO light curves, encouraging consistency between augmented views of each object. Stage~2 then combines supervised contrastive learning with an auxiliary classification loss on a separate labeled Alcock sample. This second stage uses class labels to strengthen discrimination between variability classes. The resulting encoder produces pooled representations that can be used by a separately trained classifier.

To assess what each stage contributes, we compare the representations before and after adaptation on the Alcock variable-star benchmark~\cite{donoso2026astromer2,astromer2zenodo}. We hold the original, Stage~1, and Stage~2 encoders fixed and train the same classifier architecture on their pooled representations, using 20, 100, and 500 labeled examples per class for downstream classifier training. These budgets apply to the downstream classifier; Stage~2 uses its own separate labeled sample. We also examine class organization directly in the representation space, allowing us to evaluate the learned representations beyond their performance with a particular classifier.

Our main contributions are as follows:
\begin{enumerate}
\item We introduce a two-stage contrastive adaptation of ASTROMER~2 that combines unlabeled MACHO light curves with labeled Alcock data while retaining the original encoder architecture.
\item We show that contrastive adaptation improves classification using frozen pooled representations across three downstream label budgets. On a separate 600-object holdout, the full model achieves a macro-F1 of 0.7943 with 500 labeled examples per class for downstream classifier training, exceeding our reproduction of the ASTROMER~2 finetune-and-classify pipeline by 5.10 percentage points.
\item We show through direct embedding-space diagnostics that supervised adaptation improves local class consistency and global class separation in the pooled representations.
\end{enumerate}

The remainder of this paper is organized as follows. Section~\ref{sec:background_related_work} reviews the relevant background and related work. Section~\ref{sec:contrastive_astromer} presents Contrastive Astromer, and Sections~\ref{sec:experiments} and~\ref{sec:results} describe the experiments and results. Section~\ref{sec:discussion} discusses the implications and limitations, and Section~\ref{sec:conclusion} concludes.

\section{Background and Related Work}
\label{sec:background_related_work}

\subsection{ASTROMER and Masked Reconstruction}
\label{subsec:related_foundation_light_curves}
\label{subsec:related_masked_reconstruction}

ASTROMER and ASTROMER~2 use Transformer encoders to learn representations of irregular light curves through masked-observation prediction: parts of a light curve are hidden, and the encoder predicts their magnitudes from the surrounding temporal context~\cite{donoso2023astromer,donoso2026astromer2}. ASTROMER~2 uses a deeper Transformer on fixed windows of 200 observations and is pretrained on approximately 1.5~million single-band MACHO light curves~\cite{donoso2026astromer2,astromer2zenodo}.

Masked reconstruction encourages the encoder to capture temporal context but does not explicitly require a pooled representation to separate variable-star classes. Here, a pooled representation is a single fixed-length summary vector formed by taking an equally weighted mean of the encoder's output vectors for valid observation tokens within a 200-position input window. Information useful for predicting missing magnitudes need not make class distinctions readily accessible in this summary.

\subsection{Contrastive Representation Learning}
\label{subsec:related_contrastive_learning}

Contrastive learning shapes representations through comparisons between examples. In instance-level contrastive learning, two augmented views of the same input form a positive pair, while views from other inputs in the minibatch serve as negatives. Each view acts in turn as an \emph{anchor}, or reference view. InfoNCE encourages the anchor to be more similar to its positive than to the competing negative views~\cite{oord2018representation}. SimCLR uses this pairing scheme with a normalized, temperature-scaled contrastive loss and a projection head that is discarded for downstream transfer~\cite{chen2020simple}.

The choice of augmented views determines which variations the representation is encouraged to ignore. For light curves, these variations may include small time shifts, changes in sampling, and photometric perturbations. The views should preserve the variability information relevant to the task; otherwise, the model may learn to ignore features needed for classification. Augmentation therefore defines what consistency between two views should mean.

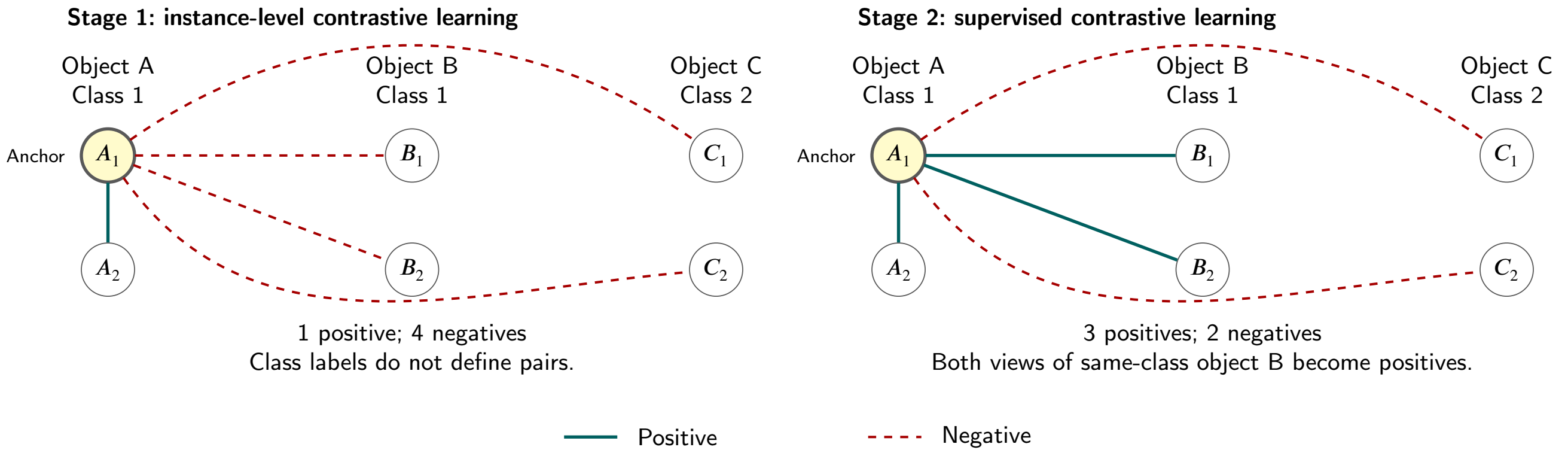
\begin{figure*}[pos=!tp]
\centering
\resizebox{\linewidth}{!}{%
\begin{tikzpicture}[
  view/.style={circle,draw=black!65,fill=white,minimum size=7mm,inner sep=0pt,font=\small},
  anchorview/.style={view,very thick,fill=yellow!25},
  positive/.style={draw=teal!75!black,line width=1.2pt},
  negative/.style={draw=red!65!black,line width=0.9pt,dashed},
  every node/.style={font=\small}
]
\foreach \stage/\offset in {1/0,2/10.4}{
  \begin{scope}[xshift=\offset cm]
    \ifnum\stage=1
      \node[anchor=west,font=\bfseries] at (-0.65,1.8) {Stage 1: instance-level contrastive learning};
    \else
      \node[anchor=west,font=\bfseries] at (-0.65,1.8) {Stage 2: supervised contrastive learning};
    \fi
    \foreach \obj/\x/\cls in {A/0/1,B/4/1,C/8/2}{
      \node[align=center] at (\x,1.0) {Object \obj\\Class \cls};
      \node[view] (\obj\stage a) at (\x,0) {$\obj_1$};
      \node[view] (\obj\stage b) at (\x,-1.5) {$\obj_2$};
    }
    \node[anchorview] at (0,0) {$A_1$};
    \node[anchor=east,font=\scriptsize] at (-0.45,0) {Anchor};
    \draw[positive] (A\stage a) -- (A\stage b);
    \ifnum\stage=1
      \draw[negative] (A\stage a) -- (B\stage a);
      \draw[negative] (A\stage a) -- (B\stage b);
      \node[align=center] at (4,-2.55) {1 positive; 4 negatives\\Class labels do not define pairs.};
    \else
      \draw[positive] (A\stage a) -- (B\stage a);
      \draw[positive] (A\stage a) -- (B\stage b);
      \node[align=center] at (4,-2.55) {3 positives; 2 negatives\\Both views of same-class object B become positives.};
    \fi
    \draw[negative] (A\stage a) to[out=35,in=145] (C\stage a);
    \draw[negative] (A\stage a) to[out=-55,in=185] (C\stage b);
  \end{scope}
}
\draw[positive] (6,-3.7) -- (6.7,-3.7);
\node[anchor=west] at (6.85,-3.7) {Positive};
\draw[negative] (10,-3.7) -- (10.7,-3.7);
\node[anchor=west] at (10.85,-3.7) {Negative};
\end{tikzpicture}
}
\caption{Positive and negative views for one anchor in a three-object minibatch. For each object, one sampled light-curve window supplies two independently augmented views, represented here by their projected embeddings (subscripts 1 and 2). Objects A and B share a class; C belongs to a different class. In Stage 1, class labels are not used: only the companion view $A_2$ is positive, while both views of B are negatives because B is a different object, despite sharing A's class. In Stage 2, class labels define the positive set, so both views of B become additional positives. The anchor itself is excluded from both sets. Only comparisons involving anchor $A_1$ are drawn; every view serves as an anchor in turn. The objects illustrate the pairing rules, not shared training data between stages.}
\label{fig:contrastive_pairs}
\end{figure*}

Class labels allow positive pairs to extend beyond two views of the same object. Without labels, instance-level learning treats different objects as negatives even when they belong to the same variability class. Supervised contrastive learning instead includes other views from the anchor's class in its positive set, while views from different classes remain negatives~\cite{khosla2020supervised}. This change encourages class-level organization in the representation space. Figure~\ref{fig:contrastive_pairs} illustrates the two pairing rules using a simple three-object minibatch.

\subsection{Representation Learning in Astronomy}
\label{subsec:related_contrastive_astronomy}

Contrastive learning has already been applied directly to astronomical light curves. StarCLR constructs positive pairs from partially overlapping subsequences of the same light curve and pretrains on TESS data~\cite{ding2026starclr}. Its use of temporal subsequences illustrates how related views can be constructed from the observations themselves, connecting the general contrastive framework to the structure of astronomical time series.

Other astronomical representation models draw on multiple photometric bands or complementary types of observation. Astro-MoE uses a Transformer with a mixture-of-experts module to model multiband astronomical time series~\cite{cadizleyton2025astromoe}. Among multimodal approaches, AstroCLIP aligns galaxy images and spectra in a shared embedding space, while AstroM$^{3}$ integrates light curves, spectra, and astrophysical metadata~\cite{lanusse2023astroclip,parker2024astroclip,rizhko2025astrom3}. Maven uses contrastive learning to align supernova photometry and spectroscopy~\cite{zhang2024maven}.

Contrastive Astromer focuses on adapting representations learned from single-band light curves. We retain the ASTROMER~2 encoder architecture and examine how unsupervised contrastive learning followed by supervised adaptation changes the usefulness and class organization of its pooled representations. This setting complements work that broadens the observational inputs by studying what can be gained from adapting an existing single-band encoder.

\section{Methodology}
\label{sec:contrastive_astromer}

\subsection{Overview}
\label{subsec:contrastive_astromer_overview}

Contrastive Astromer is a representation-learning extension of the official ASTROMER~2 light-curve Transformer. It learns a reusable light-curve representation by continuing to train a pretrained ASTROMER~2 encoder in two stages. Stage~1 uses unlabeled MACHO light curves to encourage consistency between augmented views of the same object. Stage~2 uses labeled Alcock light curves to strengthen discrimination between variability classes. 

\begin{figure*}[pos=!tp]
\centering
\includegraphics[width=\linewidth]{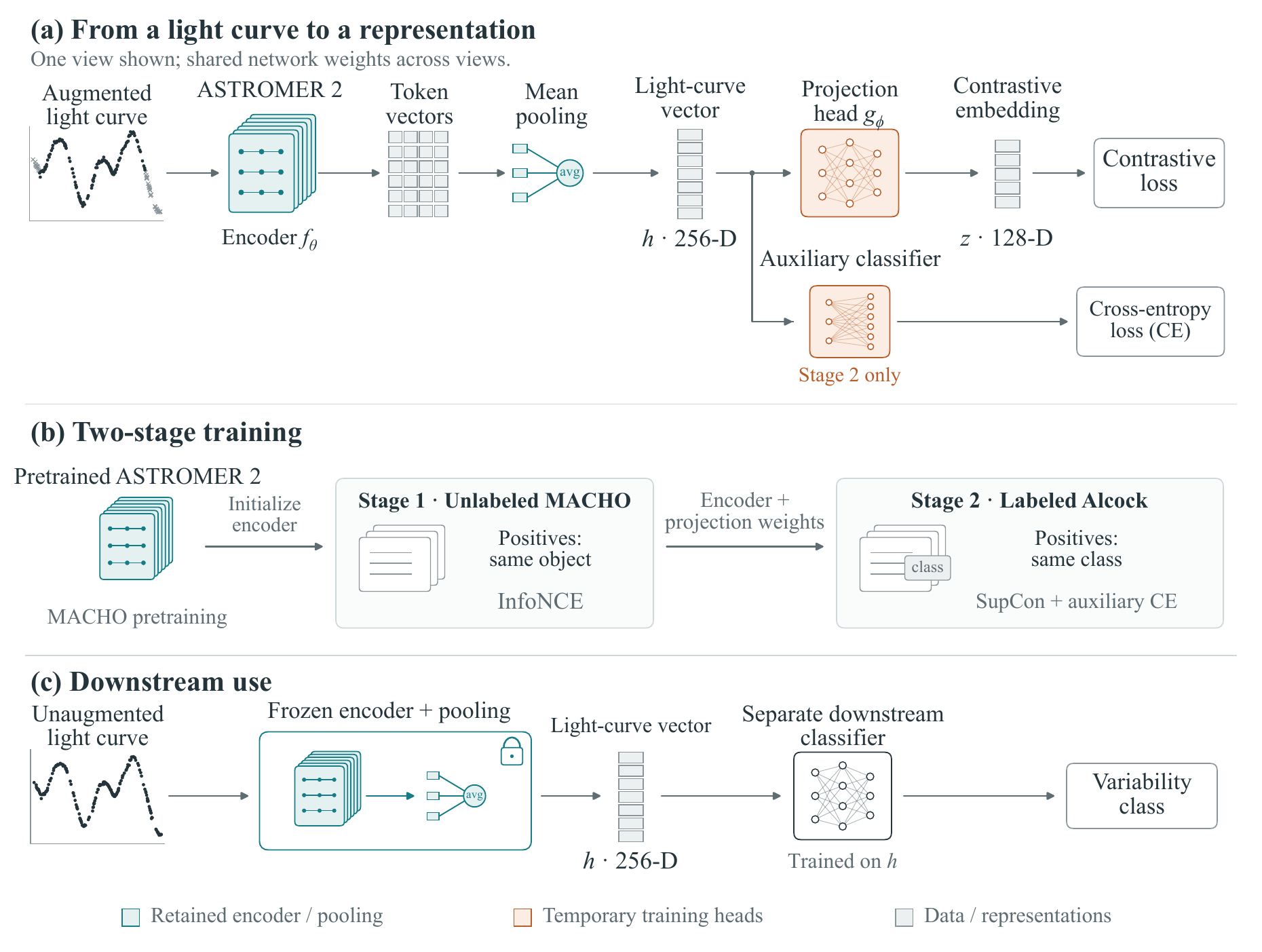}
\caption{Contrastive Astromer methodology.
(a) Each augmented view passes separately through the shared ASTROMER~2
encoder \(f_\theta\). Mean pooling over retained observations converts
the token vectors into a 256-dimensional light-curve representation \(h\).
The projection head \(g_\phi\) maps \(h\) to the normalized
128-dimensional embedding \(z\) used for contrastive learning.
The auxiliary linear classifier operates directly on \(h\) and is used
only in Stage~2.
(b) Stage~1 learns from unlabeled MACHO light curves using InfoNCE.
Its encoder and projection-head weights initialize Stage~2, which uses
labeled Alcock light curves with supervised contrastive learning and
auxiliary cross-entropy (CE).
(c) After Stage~2, both training heads are discarded. A separate downstream
classifier is trained on representations from unaugmented windows, with
the encoder frozen.
Teal denotes retained encoder and pooling components; orange denotes
temporary training heads; gray denotes data and representations.
Light-curve thumbnails show real observations; the network, token, and
vector graphics are schematic.}
\label{fig:contrastive_astromer_pipeline}
\end{figure*}

Both stages follow the same path from observations to a contrastive embedding, as illustrated in Figure~\ref{fig:contrastive_astromer_pipeline}. A light-curve window is transformed into two augmented views. For each view, the encoder produces a vector at every input position, and averaging the vectors for retained observations gives a fixed-length representation \(h\). A projection head then maps \(h\) to a normalized embedding \(z\), on which the contrastive loss is computed. Stage~2 also trains a temporary classifier on \(h\), providing an additional class-prediction signal. Figure~\ref{fig:contrastive_pairs} illustrates how the positive and negative pairs differ between stages.

The two stages change the training objective without introducing a reconstruction task. Although the initial ASTROMER~2 checkpoint was learned through masked reconstruction, neither contrastive stage predicts hidden magnitudes. Observation masking is used only to construct augmented views, while padding masks identify positions that do not contain observations. After training, we retain the encoder and pooling operation; the projection head and temporary classifier are discarded.

\subsection{Input Representation and Base Encoder}
\label{subsec:input_encoder}

Each encoder input represents a window containing at most 200 observations from one light curve. We use \(T=200\) input positions and write the magnitude and time arrays as
\begin{equation}
X = \{(m_t,\tau_t)\}_{t=1}^{T},
\end{equation}
where \(m_t\) is the magnitude and \(\tau_t\) is the observation time. A window containing \(n<T\) observations is extended with \(T-n\) padding positions, whose magnitude and time values are set to zero.

A separate binary indicator \(q_t\) records which positions are used in the current view. We write \(\mathbf{q}=(q_1,\ldots,q_T)\), with \(q_t=1\) for a retained observation and \(q_t=0\) for padding or an observation excluded by augmentation. Thus, \(q_t\) is an auxiliary mask, not an additional measured property of the light curve. It is supplied to the encoder's attention mechanism and used in pooling. Our implementation converts \(q_t\) to \(1-q_t\) to match the mask convention expected by the official encoder. During inference, every real observation in the selected window has \(q_t=1\); only padding positions have \(q_t=0\).

The base model is the six-layer ASTROMER~2 Transformer, initialized from its official MACHO-pretrained checkpoint and producing 256-dimensional latent representations~\cite{donoso2026astromer2,astromer2zenodo}. We retain its architecture, time encoding, and zero-mean normalization. The contrastive and pooled-representation pipelines read the stored magnitude values. During representation learning, the encoder weights are updated; during downstream classifier training and evaluation, they are held fixed.

Window selection determines which observations from a longer light curve are presented to the encoder. During Stage~1 and Stage~2 training, a light curve with \(L>200\) observations supplies a randomly selected contiguous window of 200 observations. The start is sampled uniformly from the valid positions. If \(L\leq200\), all observations are retained. Two augmented views are constructed from this same source window, and each view is then padded to \(T=200\) positions if necessary.

Downstream classifier training and evaluation use one deterministic window per object. For \(L>200\), we select the central 200 observations, starting at
\[
s=\left\lfloor\frac{L-200}{2}\right\rfloor
\]
with zero-based indexing. The selected window contains observations \(s\) through \(s+199\). For \(L\leq200\), we retain the full sequence and pad it. No stochastic augmentation is applied. This rule gives the original, Stage~1, and Stage~2 encoders the same observations when their pooled representations are compared. The central crop is a controlled evaluation choice rather than a claim that it is optimal.

\subsection{Pooled Encoder Representation}
\label{subsec:pooled_representation}

Mean pooling converts the encoder's sequence of outputs into one vector for the input window. Let \(f_{\theta}\) denote the ASTROMER 2 encoder with parameters \(\theta\). Given an input window \(X\), the encoder returns a sequence of token-level hidden states, denoted by \(H\), with hidden dimension \(d=256\):

\begin{equation}
H = f_{\theta}(X) = (H_1,\ldots,H_T), \qquad H_t\in\mathbb{R}^{d}.
\end{equation}

For a view containing at least one retained observation, its pooled representation is

\begin{equation}
h = \frac{\sum_{t=1}^{T} q_t H_t}{\sum_{t=1}^{T} q_t}.
\label{eq:masked_mean_pooling}
\end{equation}

The numerator sums the token vectors for retained observations, and the denominator counts those observations. Each therefore contributes equally to the resulting vector \(h\in\mathbb{R}^{256}\).

The mask selects which vectors enter the average; it does not define a reconstruction target. During contrastive training, the average excludes padding and observations removed by augmentation. At inference, it excludes only padding. If all 200 positions contain observations, Equation~\eqref{eq:masked_mean_pooling} reduces to an ordinary mean over the 200 token vectors. The pooled representation \(h\) supplies both the training heads and the separate downstream classifier.

\subsection{Contrastive Projection Head}
\label{subsec:projection_head}

The projection head provides a space specifically for the contrastive objective. Following the design used in SimCLR, we compute the contrastive loss after projection and retain the representation before projection for downstream tasks~\cite{chen2020simple}. We write
\[
z=g_\phi(h)\in\mathbb{R}^{128},
\]
where \(g_\phi\) consists of a fully connected layer with 256 units and a ReLU activation, a linear layer with 128 outputs, and \(\ell_2\) normalization.

The projection head and encoder are learned together through the contrastive loss. The head is randomly initialized at the start of Stage~1, and its weights are carried forward into Stage~2. Normalization makes the inner product between projected embeddings equivalent to cosine similarity, so comparisons depend on their directions rather than their magnitudes. Gradients through the projection head update the encoder, but downstream classification uses \(h\) directly and does not require \(g_\phi\).

\subsection{Augmentation and Contrastive Views}
\label{subsec:augmentations_views}

\begin{figure*}[pos=!tp]
\centering
\includegraphics[width=\linewidth]{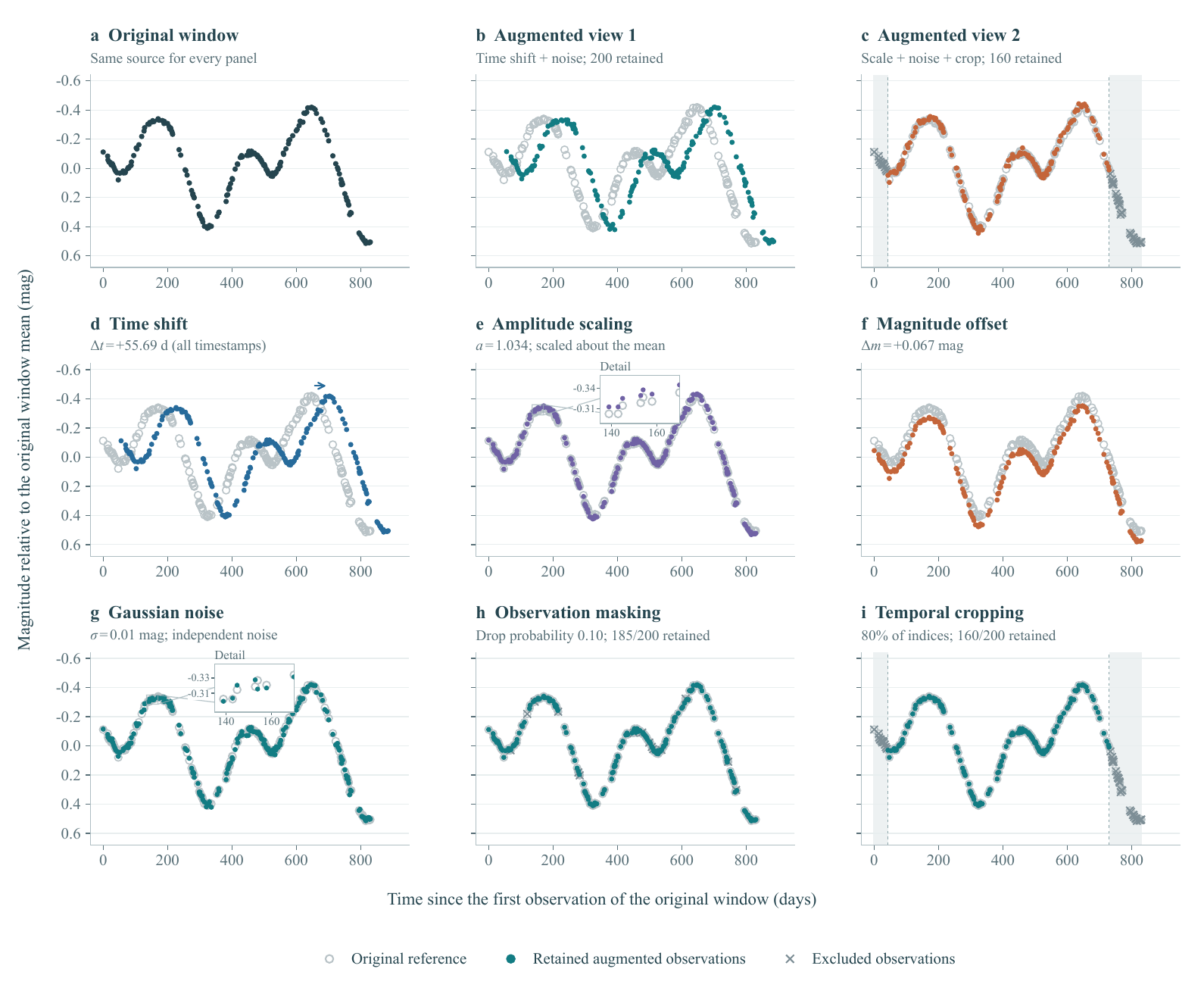}
\caption{Augmented views of a real MACHO light-curve window. Panel (a) shows 200 consecutive observations of the long-period variable MACHO 23.3302.11. Panels (b)--(c) show two independently sampled augmented views, using the Stage~1 probabilities and strengths in Table~\ref{tab:augmentation_settings}. Panels (d)--(i) illustrate each augmentation applied individually. Open gray circles show the original observations for reference, colored points show retained augmented observations, and crosses mark excluded observations. Temporal cropping retains a contiguous interval of observation indices; masking and cropping change visibility without introducing zero-valued measurements. All main panels share the same axes, original time origin, and original-window mean magnitude, which is subtracted for display only. Insets magnify the small changes from amplitude scaling and Gaussian noise without changing their strength.}
\label{fig:augmented_light_curves_complex}
\end{figure*}

Augmentation creates two related views of each source window for contrastive training. For a sampled window \(x_i\), we independently draw
\begin{equation}
x_i^{(1)},x_i^{(2)}\sim\mathcal{T}(x_i),
\end{equation}
where \(\mathcal{T}\) denotes the augmentation distribution. Each view contains its own magnitude array, time array, and observation mask. Only these two views enter the contrastive objective; the unaugmented source window is not included as a third view. Figure~\ref{fig:contrastive_pairs} shows how the resulting views form positive and negative pairs. Figure~\ref{fig:augmented_light_curves_complex} illustrates this construction for a real light-curve window, showing the original observations, two independently sampled views, and the effects of the six individual augmentation operators.

The value-based augmentations perturb the measurements while retaining their array positions. A random time shift adds the same offset to all timestamps in a view, with the offset bounded by a fraction of the window's time span. Amplitude scaling rescales magnitude deviations around the window mean, and a magnitude offset adds a constant to all magnitudes. Gaussian noise adds independent, zero-mean perturbations to the magnitude values. These transformations are intended to preserve the variability information relevant to classification.

Observation masking and temporal cropping instead change which positions are retained. Random masking independently excludes observations by setting their \(q_t\) values to zero. Temporal cropping selects a contiguous interval of observation indices and sets \(q_t=0\) outside that interval. The crop fraction refers to the number of positions retained within the interval, rather than a fraction of elapsed time. Neither operation deletes array entries: excluded positions are omitted as attention keys and from pooling.

Stage~2 uses the same augmentation family with lower application probabilities and smaller perturbations. Table~\ref{tab:augmentation_settings} reports the settings for both stages.

\begin{table*}[pos=!tp]
\centering
\caption{Augmentation probabilities and strengths for Stage 1 and Stage 2.}
\label{tab:augmentation_settings}
\small
\setlength{\tabcolsep}{3pt}
\renewcommand{\arraystretch}{1.18}
\begin{tabular}{@{}>{\raggedright\arraybackslash}p{0.20\linewidth}>{\centering\arraybackslash}p{0.145\linewidth}>{\raggedright\arraybackslash}p{0.22\linewidth}>{\centering\arraybackslash}p{0.145\linewidth}>{\raggedright\arraybackslash}p{0.22\linewidth}@{}}
\toprule
& \multicolumn{2}{c}{Stage 1} & \multicolumn{2}{c}{Stage 2} \\
\cmidrule(lr){2-3}\cmidrule(l){4-5}
Augmentation operator & Probability & Strength / parameter & Probability & Strength / parameter \\
\midrule
Time jitter & \(0.5\) & Maximum shift: \(0.1\) times time range & \(0.2\) & Maximum shift: \(0.03\) times time range \\
\addlinespace
Amplitude scaling & \(0.5\) & Scale range: \([0.95,1.05]\) & \(0.25\) & Scale range: \([0.98,1.02]\) \\
\addlinespace
Amplitude shift & \(0.3\) & Shift range: \([-0.1,0.1]\) & \(0.1\) & Shift range: \([-0.03,0.03]\) \\
\addlinespace
Gaussian magnitude noise & \(0.8\) & Noise level: \(0.01\) & \(0.35\) & Noise level: \(0.005\) \\
\addlinespace
Random observation masking & \(0.3\) & Drop probability: \(0.1\) & \(0.1\) & Drop probability: \(0.05\) \\
\addlinespace
Temporal cropping & \(0.2\) & Crop fraction: \(0.8\) & \(0.05\) & Crop fraction: \(0.9\) \\
\bottomrule
\end{tabular}
\par\smallskip
\begin{minipage}{\linewidth}
\footnotesize
Probability is the chance of applying the augmentation to a view. For observation masking, the separate drop probability applies to each currently visible observation when masking is applied.
\end{minipage}
\end{table*}


\subsection{Stage 1: Unsupervised Contrastive Learning on MACHO}
\label{subsec:stage1}

Stage~1 trains the encoder to recognize two augmented views of the same source window. It starts from the MACHO-pretrained ASTROMER~2 encoder and a randomly initialized projection head. A minibatch of \(N\) source windows produces \(2N\) normalized projected embeddings, which we index as \(\{z_a\}_{a=1}^{2N}\). Each view acts once as an anchor, and its companion view from the same window is its positive. Both views of every other object serve as negatives.

We optimize the normalized temperature-scaled contrastive loss used in SimCLR, an InfoNCE-style objective~\cite{oord2018representation,chen2020simple}. For anchor \(a\), let \(p(a)\) index its positive view. The loss is

\begin{equation}
\ell_a =
-\log
\frac{
\exp\left(\operatorname{sim}(z_a,z_{p(a)})/\tau\right)
}{
\sum_{b\neq a}
\exp\left(\operatorname{sim}(z_a,z_b)/\tau\right)
},
\label{eq:nt_xent_single}
\end{equation}

Here, \(\operatorname{sim}(z_a,z_b)=z_a^\top z_b\), since the projected embeddings are normalized. The denominator includes the positive and all \(2N-2\) negatives; only the anchor itself is excluded.

The temperature \(\tau\) controls how strongly the loss emphasizes high-similarity comparisons. Smaller values place greater weight on the most similar negatives, which are hardest to distinguish from the positive. Larger values distribute that weighting more evenly across the competing views.

The Stage~1 objective averages the loss over all anchors:
\begin{equation}
\mathcal{L}_{\mathrm{stage1}}
=
\mathcal{L}_{\mathrm{InfoNCE}}
=
\frac{1}{2N}\sum_{a=1}^{2N}\ell_a.
\label{eq:stage1_loss}
\end{equation}
This objective updates both the encoder and projection head. No class labels or magnitude-reconstruction targets are used. The training settings and checkpoint-selection procedure are described in Section~\ref{sec:experiments}.

\subsection{Stage 2: Supervised Contrastive Learning on Alcock}
\label{subsec:stage2}

Stage~2 uses class labels to extend the positive set beyond another view of the same object. It loads both the encoder and projection-head weights from a Stage~1 checkpoint and continues training on labeled Alcock light curves. The six classes are \texttt{Cep\_0}, \texttt{Cep\_1}, \texttt{EC}, \texttt{LPV}, \texttt{RRab}, and \texttt{RRc}.

Each labeled object again supplies two augmented views. Let \(y_a\) denote the class label inherited by view \(a\). Its positive set contains every other view in the minibatch with the same label:

\begin{equation}
P(a)=\{p\in\{1,\ldots,2N\}: p\neq a,\; y_p=y_a\}.
\label{eq:supcon_positive_set}
\end{equation}

This set includes the companion view of the same object and, when present, views of other objects in the same class. Every anchor therefore has at least one positive, so the set of valid anchors is
\begin{equation}
\mathcal{I}=\{1,\ldots,2N\}.
\label{eq:supcon_valid_anchors}
\end{equation}

The supervised contrastive objective averages the log-probability of the same-class positives for each anchor:

\begin{equation}
\begin{aligned}
\mathcal{L}_{\mathrm{supcon}}
&=
\frac{1}{|\mathcal{I}|}
\sum_{a\in\mathcal{I}}
\\
&\quad
\left[
\frac{-1}{|P(a)|}
\sum_{p\in P(a)}
\log
\frac{
\exp\left(\operatorname{sim}(z_a,z_p)/\tau\right)
}{
\sum_{b\neq a}
\exp\left(\operatorname{sim}(z_a,z_b)/\tau\right)
}
\right].
\end{aligned}
\label{eq:supcon_loss}
\end{equation}

The denominator contains all non-anchor views, including every positive. Views with different class labels serve as negatives.

An auxiliary classifier provides a class-prediction signal directly to the representation retained for downstream use. We attach a randomly initialized linear classifier \(c_\psi\) to \(h\), with six output logits. Its cross-entropy loss is averaged over both augmented views of every object:
\begin{equation}
\mathcal{L}_{\mathrm{ce}}
=
-\frac{1}{2N}
\sum_{a=1}^{2N}
\log\!\left[
\operatorname{softmax}\!\left(c_\psi(h_a)\right)_{y_a}
\right].
\label{eq:ce_loss}
\end{equation}
Thus, supervised contrastive learning acts on \(z\), while the auxiliary classifier encourages class discrimination directly through \(h\). This classifier is used only during Stage~2 and is separate from the downstream classifier.

The two objectives are combined with fixed weights that also specify the training schedule:
\begin{equation}
\mathcal{L}_{\mathrm{stage2}}
=
\lambda_{\mathrm{sup}}\mathcal{L}_{\mathrm{supcon}}
+
\lambda_{\mathrm{ce}}\mathcal{L}_{\mathrm{ce}}.
\label{eq:stage2_loss}
\end{equation}
The first epoch uses a cross-entropy warm-up,
\[
(\lambda_{\mathrm{sup}},\lambda_{\mathrm{ce}})=(0,1),
\]
which updates the encoder and the new auxiliary classifier. During epochs 2--15, we use
\[
(\lambda_{\mathrm{sup}},\lambda_{\mathrm{ce}})=(1,0.1),
\]
so both objectives update the encoder, with the projection head trained through the contrastive term and the auxiliary classifier trained through cross-entropy.

Section~\ref{sec:experiments} describes the data partitions, optimization settings, and selection of the final encoder.

\subsection{Final Contrastive Astromer Encoder}
\label{subsec:final_contrastive_astromer_encoder}

The final model produces one fixed-length representation from an unaugmented light-curve window. After Stage~2, we discard the projection head and auxiliary classifier, retaining the trained encoder and the mean-pooling operation in Equation~\eqref{eq:masked_mean_pooling}.

For downstream use, we select the deterministic window described in Section~\ref{subsec:input_encoder}, retain every real observation, and mark only padding positions with \(q_t=0\). The encoder remains frozen while a separate classifier is trained on \(h\), as described in Section~\ref{subsec:downstream_evaluation}.

\section{Experiments}
\label{sec:experiments}

This section describes the data partitions, training settings, and validation-only selection procedure used before final holdout evaluation.

\subsection{Datasets}
\label{subsec:experiments_datasets}

We use the MACHO and Alcock datasets distributed through the ASTROMER~2 GitHub repository\footnote{\url{https://github.com/astromer-science/main-code}}~\cite{donoso2026astromer2}. To accommodate our two-stage setup and prevent data leakage between stages, we repartition the data into object-disjoint pools for Stage 1, Stage 2, downstream cross-validation, and final testing.

We initialize Stage 1 from the MACHO-pretrained ASTROMER~2 checkpoint for unsupervised contrastive learning. Because the MACHO data share objects with the labeled Alcock catalog, we first remove all overlapping objects from the Stage 1 training set. The resulting Stage 1 training set shares no objects with Stage 2, downstream cross-validation, or the final holdout.

The Alcock catalog contains 20,894 labeled objects spanning six classes: \texttt{Cep\_0}, \texttt{Cep\_1}, \texttt{EC}, \texttt{LPV}, \texttt{RRab}, and \texttt{RRc}. We use a subset of 17,294 objects for supervised contrastive adaptation in Stage 2. Class counts in this subset are unequal. The remaining 3,600 objects are split into a 3,000-object downstream pool (500 per class) and a separate 600-object final holdout (100 per class). The Alcock20, Alcock100, and Alcock500 settings draw 20, 100, and 500 objects per class from the downstream pool, respectively, for totals of 120, 600, and 3,000 objects. Each setting is evaluated with five non-overlapping validation folds, giving train/validation counts per class of 16/4, 80/20, and 400/100, respectively. The same folds are used for the pooled-\(h\) controls and the reproduced official ASTROMER~2 pipeline. The final holdout is excluded from training, validation, and encoder selection. Table~\ref{tab:alcock_split_sizes} summarizes the mutually disjoint Alcock partitions.

\begin{table*}[pos=!tp]
\centering
\small
\caption{Object-disjoint Alcock partitions used in the experiments. Counts denote unique objects.}
\label{tab:alcock_split_sizes}
\begin{tabular}{lrrl}
\toprule
Partition & Objects & Per class & Role \\
\midrule
Stage 2 pool & 17,294 & class-imbalanced & representation learning \\
Downstream CV pool & 3,000 & 500 & five-fold selection \\
Final holdout & 600 & 100 & one-time testing \\
\midrule
Total & 20,894 & --- & --- \\
\bottomrule
\end{tabular}
\par\smallskip
\begin{minipage}{\linewidth}
\footnotesize
The downstream pool contains 500 objects per class. Alcock20, Alcock100, and Alcock500 use 20, 100, and 500 objects per class from this pool (120, 600, and 3,000 objects in total), respectively. The final holdout is separate and contains 100 objects per class.
\end{minipage}
\end{table*}

\subsection{Training and Evaluation Setup}
\label{subsec:training_protocol}

Table~\ref{tab:experiment_protocol} separates encoder construction, validation-only encoder selection, and final evaluation. The encoder is updated only in Stage 1 and Stage 2. Downstream classification freezes the selected encoder and trains a new classifier on its pooled representation \(h\).

\begin{table*}[pos=!tp]
\centering
\scriptsize
\setlength{\tabcolsep}{2pt}
\renewcommand{\arraystretch}{1.12}
\caption{Data roles in the Contrastive Astromer workflow.}
\label{tab:experiment_protocol}
\begin{tabular}{p{0.15\linewidth}>{\raggedright\arraybackslash}p{0.20\linewidth}>{\raggedright\arraybackslash}p{0.09\linewidth}>{\raggedright\arraybackslash}p{0.10\linewidth}>{\raggedright\arraybackslash}p{0.21\linewidth}>{\raggedright\arraybackslash}p{0.15\linewidth}}
\toprule
Stage & Dataset and split & Labels used? & Encoder updated? & Projection/\newline classifier head & Purpose \\
\midrule
Stage 1 & MACHO; no objects overlapping with Alcock & No & Yes & Projection head trained & Unsupervised contrastive representation learning \\
\addlinespace[4pt]
Stage 2 & 17,294-object Alcock representation pool & Yes & Yes & Projection head and auxiliary classifier trained & Supervised contrastive learning \\
\addlinespace[4pt]
CV and\newline selection & 3,000-object pool; five folds; Alcock20/100/500 & Yes & No & Pooled-\(h\) MLP head trained per run & Select one Stage 1 checkpoint using validation only \\
\addlinespace[4pt]
Final\newline evaluation & Setting-specific training pool; 600-object holdout & Yes & No & Final MLP head retrained & Report holdout performance \\
\bottomrule
\end{tabular}
\end{table*}

The encoder architecture remains unchanged throughout. Stage 1 begins from the official MACHO-pretrained ASTROMER~2 checkpoint. Stage 1 runs for 100 epochs, saving a checkpoint after every epoch. Each checkpoint initializes a separate 15-epoch Stage 2 run on the disjoint Alcock subset. Table~\ref{tab:contrastive_training_settings} summarizes the training settings.

\begin{table*}[pos=!tp]
\centering
\small
\setlength{\tabcolsep}{5pt}
\renewcommand{\arraystretch}{1.2}
\caption{Training and evaluation settings for the two encoder-construction stages and the downstream classifier head.}
\label{tab:contrastive_training_settings}
\begin{tabular}{@{}>{\raggedright\arraybackslash}p{0.17\linewidth}>{\raggedright\arraybackslash}p{0.24\linewidth}>{\raggedright\arraybackslash}p{0.27\linewidth}>{\raggedright\arraybackslash}p{0.22\linewidth}@{}}
\toprule
Setting & Stage 1 \newline (unsupervised contrastive) & Stage 2 \newline (supervised contrastive) & Downstream \newline classifier head \\
\midrule
Data split & MACHO; no objects overlapping with Alcock & 17,294-object Alcock pool & 3,000-object CV pool; 600-object holdout \\
Input window & 200 observations & 200 observations & central 200-observation window \\
Encoder & updated & updated & frozen \\
Objective & InfoNCE / NT-Xent & supervised contrastive + auxiliary CE & sparse categorical cross-entropy \\
Optimizer & Adam & Adam & Adam \\
Learning rate & \(10^{-4}\) & \(10^{-4}\) & \(10^{-4}\) \\
Batch size & 32 & 32 & 128 \\
Temperature \(\tau\) & 0.07 & 0.07 & --- \\
Augmentation & main (Table~\ref{tab:augmentation_settings}) & weak (Table~\ref{tab:augmentation_settings}) & none \\
Epochs / schedule & 100 & 15; epoch 1 CE warm-up \((\lambda_{\mathrm{sup}},\lambda_{\mathrm{ce}})=(0,1)\), epochs 2--15 \((1,0.1)\) & early stopping, patience 40 \\
Checkpoint use & saved every epoch & epoch-15 encoder used for each Stage 1 checkpoint & best validation-loss head restored within each fold \\
Trained head & projection head & projection head + auxiliary classifier on \(h\) & MLP \(1024\)--\(512\)--\(256\) + LayerNorm, 6-class output \\
\bottomrule
\end{tabular}
\end{table*}

Stage~2 follows the loss schedule defined in Section~\ref{subsec:stage2}. The first epoch optimizes only the auxiliary classifier's cross-entropy loss, \(\mathcal{L}_{\mathrm{ce}}\), defined in Equation~\eqref{eq:ce_loss}, with \((\lambda_{\mathrm{sup}},\lambda_{\mathrm{ce}})=(0,1)\). Epochs 2--15 optimize the combined objective in Equation~\eqref{eq:stage2_loss}, with \((\lambda_{\mathrm{sup}},\lambda_{\mathrm{ce}})=(1,0.1)\). Every Stage~2 run lasts exactly 15 epochs, and downstream evaluation uses the encoder saved at the end of epoch~15. This training duration is fixed across all runs. The validation procedure described below selects among the resulting encoders initialized from different Stage~1 checkpoints.

\subsubsection{Downstream Cross-Validation and Final Evaluation}
\label{subsec:downstream_evaluation}

Downstream evaluation compares three fixed encoder representations: the original MACHO-pretrained ASTROMER~2 encoder, a Stage 1 encoder, and its corresponding Stage 2 encoder. For each light curve, we encode its central segment and apply the mean in Equation~\eqref{eq:masked_mean_pooling} to the resulting token representations to obtain the pooled representation \(h\). A separate classifier is trained on each encoder's pooled representations using the same classifier architecture, training settings, and data splits; the encoder remains frozen, and the contrastive projection head is not used.

The downstream classifier is an MLP that takes the pooled representation \(h\) as input. It has three ReLU hidden layers with 1024, 512, and 256 units, followed by layer normalization and a six-class output layer. These hidden-layer widths match those of the official ASTROMER~2 classifier. Our classifier receives only \(h\), the mean of the final Transformer layer's output vectors over the real observations in the selected window. The official classifier instead combines pooled representations from the input embedding and all Transformer layers, as described in Section~\ref{subsec:experiments_baselines}. We optimize sparse categorical cross-entropy from logits with Adam at learning rate \(10^{-4}\), batch size 128, and random seed 42. Validation loss controls early stopping with patience 40, and the minimum-validation-loss checkpoint is restored.

Each candidate encoder is evaluated through a separately trained downstream classifier. For every Alcock setting and fold, we train an MLP with the encoder frozen and retain the classifier weights from the training epoch with the lowest validation cross-entropy loss. We record the validation macro-F1 at that same epoch. Thus, the reported validation macro-F1 corresponds to the loss-selected classifier checkpoint, rather than the highest macro-F1 reached during classifier training.

We then use these validation scores to select the final encoder. Each of the 100 Stage~1 checkpoints initializes a separate Stage~2 run, whose epoch-15 encoder is evaluated across the three Alcock settings and five folds. This gives \(100\times3\times5=1{,}500\) downstream classifier runs. For each candidate encoder, we average the validation macro-F1 values from its 15 selected classifier checkpoints, giving every setting and fold equal weight. We select the encoder with the highest mean validation macro-F1; exact ties are resolved by lower mean validation loss and then by the earlier Stage~1 initialization checkpoint. Test metrics do not enter this selection. The selected Stage~2 encoder is used across all three label budgets.

We also evaluate Stage 1 without Stage 2. We compare all 100 Stage 1 checkpoints using the same validation criterion across the three Alcock settings and five folds, with the same tie-breaking rules. This comparison selects the checkpoint used for the standalone Stage 1 results. We additionally evaluate the checkpoint selected for the full pipeline before Stage 2, allowing a direct comparison of the same encoder before and after Stage 2 training.

After encoder selection, we choose the number of training epochs for each final classifier from the mean validation-loss curve across the five cross-validation folds, using only epochs observed in all five folds. We then initialize a new classifier and train it for that number of epochs on the combined training and validation objects for the corresponding Alcock setting, without further early stopping. The classifier is evaluated once on the separate 600-object holdout. For the official ASTROMER~2 baseline, the numbers of epochs for target finetuning and classifier training are likewise selected through cross-validation.

\subsection{Baselines}
\label{subsec:experiments_baselines}

We use two ASTROMER~2 baselines to compare the complete classification pipeline and the encoder representation itself. The first reproduces the official finetune-and-classify pipeline across all three Alcock settings using the same disjoint five-fold downstream data construction. Its classifier draws on representations from several depths of the encoder: the input embedding before the first Transformer layer and the output of every Transformer layer. At each depth, it averages the token vectors over non-padding observations to obtain one vector per light-curve window. It then combines these pooled vectors using learned layer weights, implemented by the \texttt{GammaWeight} layer, and passes the result to an MLP. This baseline therefore includes both the official finetuning procedure and its multilayer readout.

The second baseline evaluates the original MACHO-pretrained ASTROMER~2 encoder through the same frozen-encoder classifier used for our contrastive models. It uses only the final-layer pooled representation \(h\), with the encoder held fixed while a separate MLP is trained as described in Section~\ref{subsec:downstream_evaluation}. Comparing this baseline with the Stage~1 and Stage~2 encoders measures the changes in classification performance under a common pooling rule, classifier architecture, and downstream training protocol.

\subsection{Metrics}
\label{subsec:experiments_metrics}

The primary metric is macro-F1, the unweighted mean of the class-specific F1 scores. It gives each class equal weight, regardless of how many objects the class contains. Metric choice does not itself rebalance the training data or loss.

We also record accuracy, macro precision, macro recall, per-class precision, recall, and F1, confusion matrices, validation loss, and the selected epoch. The final holdout contains 100 objects per class, so accuracy is not dominated by unequal class support; macro-F1 remains primary because it captures the balance between precision and recall within every class. For the checkpoint-selection trajectories in Figure~\ref{fig:checkpoint_selection}, the curves show aggregate mean validation macro-F1 and the shaded bands show the sample standard deviation across the setting/fold runs at each Stage 1 epoch. Final-refit holdout metrics are reported as single values.

\subsection{Implementation Details}
\label{subsec:experiments_implementation}

The experiments are implemented in TensorFlow/Keras~\cite{abadi2015tensorflow,chollet2015keras} and retain the official ASTROMER~2 encoder architecture and checkpoint interface~\cite{donoso2026astromer2,astromer2zenodo}. The appendix records the mask convention and the data-disjointness audit.

\section{Results}
\label{sec:results}

\begin{figure*}[pos=!tp]
\centering
\includegraphics[width=\linewidth]
{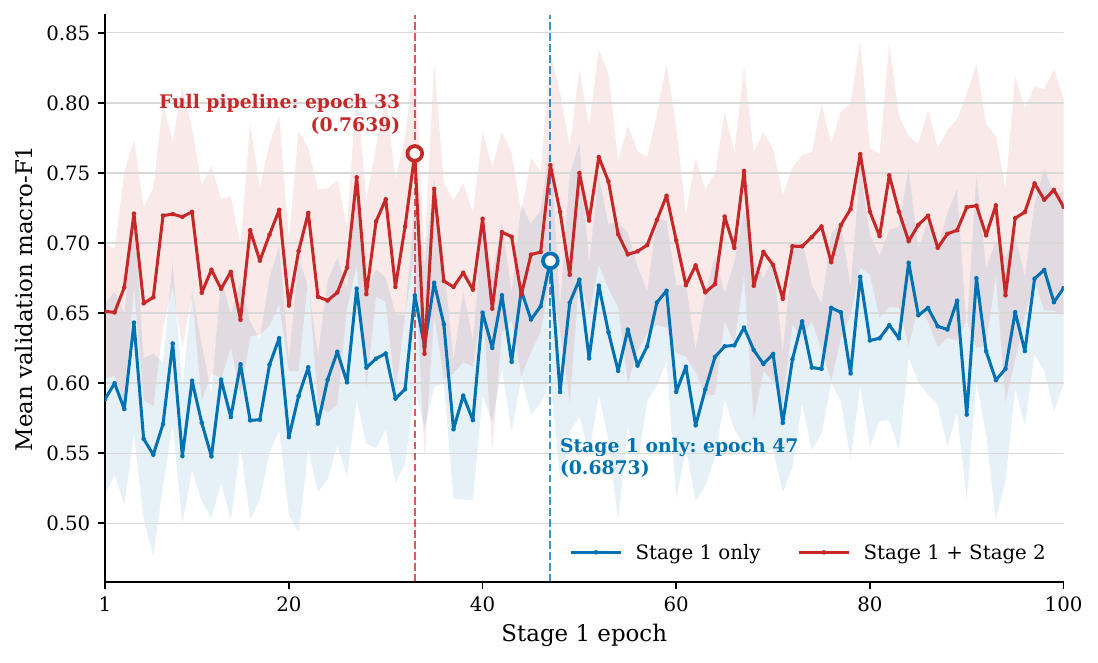}

\caption{Downstream validation macro-F1 across all 100 Stage~1 checkpoints for standalone Stage~1 and the full two-stage pipeline. The horizontal axis identifies the Stage~1 checkpoint: the standalone curve evaluates that encoder directly, while the two-stage curve evaluates the encoder obtained after a further 15 epochs of Stage~2 training. Each point averages scores across three Alcock settings and five folds; shaded bands denote the sample standard deviation across these 15 runs. Validation selects Stage~1 epoch~47 for standalone classification and epoch~33 as the initialization for the full pipeline.}
\label{fig:checkpoint_selection}
\end{figure*}

Downstream validation favors different Stage~1 checkpoints depending on whether the encoder is used directly or receives further supervised adaptation. We evaluate each Stage~1 checkpoint both on its own and after a fixed 15 epochs of Stage~2 training, fitting a separate downstream MLP to each frozen encoder. Following the selection procedure in Section~\ref{subsec:downstream_evaluation}, standalone Stage~1 achieves its highest mean validation macro-F1 at epoch~47 (\(0.6873\)), whereas the full two-stage pipeline achieves its highest score when initialized from Stage~1 epoch~33 (\(0.7639\); Figure~\ref{fig:checkpoint_selection}).

These two Stage~1 checkpoints serve distinct roles in the comparisons below. Epoch~47 represents Stage~1 selected for standalone classification. Epoch~33 provides the matched starting point for measuring the improvement introduced by Stage~2: we compare this encoder before and after the 15 epochs of supervised adaptation. Both the classification and embedding-space evaluations include this matched comparison, while the classification results also include the independently selected epoch-47 encoder.

\subsection{Downstream Classification}
\label{subsec:results_main_downstream}

Stage~2 achieves the highest test macro-F1 across all three downstream label budgets. Table~\ref{tab:main_results} reports final-refit holdout scores for the selected Stage~2 encoder, both Stage~1 controls, the original ASTROMER~2 encoder under the same pooled-\(h\) interface, and the reproduced official ASTROMER~2 pipeline. For each contrastive model, the same encoder checkpoint is used across all three label budgets.

\begin{table*}[pos=!tp]
\centering
\small
\setlength{\tabcolsep}{3pt}
\caption{Final-refit test macro-F1 after five-fold cross-validation model and training-duration selection. Each final model is refit on the complete development pool and evaluated once on the fixed holdout.}
\label{tab:main_results}
\begin{tabular}{@{}>{\raggedright\arraybackslash}p{0.285\linewidth}
                    >{\centering\arraybackslash}p{0.205\linewidth}
                    >{\centering\arraybackslash}p{0.205\linewidth}
                    >{\centering\arraybackslash}p{0.205\linewidth}@{}}
\toprule
Model / Pipeline & Alcock20 & Alcock100 & Alcock500 \\
\midrule
Official ASTROMER~2 finetune-and-classify
    & \(0.6933\) & \(0.7313\) & \(0.7433\) \\
Original ASTROMER~2 encoder pooled-\(h\)
    & \(0.1735\) & \(0.2700\) & \(0.3599\) \\
Stage 1, epoch 47 (independently selected)
    & \(0.6248\) & \(0.6801\) & \(0.7289\) \\
Stage 1, epoch 33 (matched initialization)
    & \(0.6056\) & \(0.6703\) & \(0.7429\) \\
Stage 2 (from Stage 1, epoch 33)
    & \(\mathbf{0.7078}\) & \(\mathbf{0.7704}\) & \(\mathbf{0.7943}\) \\
\bottomrule
\end{tabular}
\vspace{2pt}
\begin{minipage}{0.98\linewidth}
\footnotesize
\end{minipage}
\end{table*}

The advantage over the reproduced official pipeline grows with the downstream label budget. Stage~2 improves test macro-F1 by \(0.0145\), \(0.0391\), and \(0.0510\) on Alcock20, Alcock100, and Alcock500, respectively, with corresponding accuracies of \(0.7033\), \(0.7700\), and \(0.7933\). Performance increases with the downstream label budget for every model, and Stage~2 maintains its lead throughout.

Stage~2 also outperforms the Stage~1 encoder selected specifically for standalone classification. This epoch-47 encoder achieves final-refit test macro-F1 scores of \(0.6248\), \(0.6801\), and \(0.7289\) on Alcock20, Alcock100, and Alcock500, respectively. Its checkpoint was selected using downstream validation performance before any Stage~2 training.

The matched epoch-33 comparison reveals how each contrastive stage contributes to the final result. Under the common pooled-\(h\) interface, Stage~1 improves over the original ASTROMER~2 encoder by \(0.4321\), \(0.4003\), and \(0.3830\) macro-F1 on Alcock20, Alcock100, and Alcock500, respectively. This Stage~1 encoder remains below the complete official pipeline by \(0.0877\), \(0.0610\), and \(0.0004\), as expected given the official pipeline's additional target finetuning and different readout.

Stage~2 builds on these gains, with its largest incremental benefit in the lowest-label setting. Relative to the matched epoch-33 Stage~1 encoder, it adds \(0.1022\), \(0.1001\), and \(0.0514\) macro-F1 on Alcock20, Alcock100, and Alcock500, respectively. Supervised adaptation therefore improves classification at every downstream label budget. Because the comparison follows the same encoder before and after Stage~2, these gains measure the effect of supervised contrastive adaptation rather than per-setting checkpoint selection.

Allowing a separate Stage~1 initialization for each Alcock setting would select epoch~52 for Alcock20 and epoch~79 for Alcock100 and Alcock500. These per-setting selections are retrospective sensitivity analyses; the primary Stage~2 results use the globally selected epoch-33 initialization across all three budgets, with the same checkpoint providing the matched Stage~1 control.

Table~\ref{tab:per_class_f1_stage2_alcock500} reports the Alcock500 class-level results for the final-refit classifier built on the selected Stage~2 encoder initialized from Stage~1 epoch~33. Values are computed on the fixed 600-object holdout, which contains 100 objects per class.

\begin{table}[pos=!htbp]
\centering
\caption{Per-class Alcock500 performance for the final-refit classifier built on the globally selected Stage 2 encoder.}
\label{tab:per_class_f1_stage2_alcock500}
\setlength{\tabcolsep}{4pt}
\begin{tabular}{lcccc}
\toprule
Class & Precision & Recall & F1 & Support \\
\midrule
\texttt{Cep\_0} & 0.8977 & 0.7900 & 0.8404 & 100 \\
\texttt{Cep\_1} & 0.6484 & 0.8300 & 0.7281 & 100 \\
\texttt{EC}     & 0.7912 & 0.7200 & 0.7539 & 100 \\
\texttt{LPV}    & 0.8353 & 0.7100 & 0.7676 & 100 \\
\texttt{RRab}   & 0.8367 & 0.8200 & 0.8283 & 100 \\
\texttt{RRc}    & 0.8091 & 0.8900 & 0.8476 & 100 \\
\bottomrule
\end{tabular}
\end{table}

\begin{figure}[pos=!htbp]
\centering
\includegraphics[width=\linewidth]{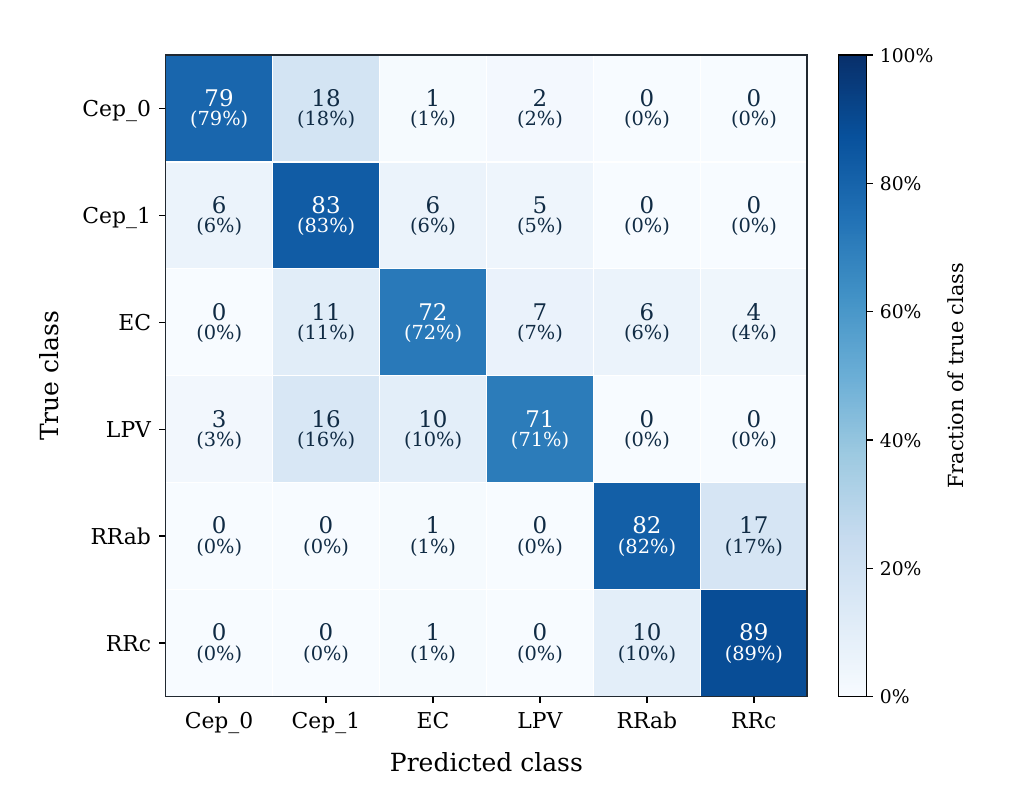}
\caption{Alcock500 confusion matrix for the final-refit classifier built on the globally selected Stage 2 encoder. Rows are true classes and columns are predictions on the fixed 600-object holdout; each row contains 100 objects.}
\label{tab:confusion_matrix_stage2_alcock500}
\end{figure}

Alcock500 F1 is lowest for \texttt{Cep\_1} (\(0.7281\)) and highest for \texttt{RRc} (\(0.8476\)), \texttt{Cep\_0} (\(0.8404\)), and \texttt{RRab} (\(0.8283\)). Figure~\ref{tab:confusion_matrix_stage2_alcock500} shows that the largest individual off-diagonal errors are \texttt{Cep\_0}\(\rightarrow\)\texttt{Cep\_1} (18), \texttt{RRab}\(\rightarrow\)\texttt{RRc} (17), and \texttt{LPV}\(\rightarrow\)\texttt{Cep\_1} (16). \texttt{RRab}/\texttt{RRc} remains the largest mutual confusion, with 17 and 10 cross-assignments.

The gains from Stage~2 extend to the Cepheid classes, which are strongly underrepresented in the adaptation data. The Stage~2 representation-learning pool contains only 73 \texttt{Cep\_1} and 572 \texttt{Cep\_0} objects, compared with \(6{,}214\) \texttt{EC} and \(6{,}787\) \texttt{RRab} objects. Despite this imbalance, Stage~2 raises Alcock500 F1 from \(0.5488\) to \(0.7281\) for \texttt{Cep\_1} and from \(0.7822\) to \(0.8404\) for \texttt{Cep\_0}, relative to the matched epoch-33 Stage~1 encoder. The final classifier correctly identifies 83 of the 100 \texttt{Cep\_1} holdout objects, achieving a recall of \(0.8300\) rather than collapsing this minority class into a majority class. These results indicate that supervised contrastive adaptation substantially strengthens the representation of difficult minority classes despite the imbalance in its adaptation data.


\subsection{Embedding-Space Evaluation}
\label{subsec:results_embedding_quality}

Downstream classification establishes whether the embeddings support accurate prediction, but it does not reveal how the classes are arranged in the representation space. We therefore examine whether contrastive training makes the embeddings themselves more class-coherent and better separated. Projection and clustering-based diagnostics are established ways to inspect the structure of learned representation spaces~\cite{xie2016deep,mitrovic2020less}. We first inspect the 256-dimensional pooled representations \(h\) visually. PCA, t-SNE, and UMAP reduce the embeddings to two dimensions without using the class labels~\cite{vandermaaten2008visualizing,mcinnes2018umap}. After each projection is computed, we color the points by their known classes solely to make class structure visible.

The visualization uses a class-balanced sample of 3,000 objects drawn from the complete 20,894-object Alcock pool, with 500 objects sampled from each class. The exact sampling, preprocessing, and projection settings are reported in Appendix~\ref{app:projection_settings}.

Figure~\ref{fig:embedding_projection_grid} compares the original ASTROMER~2, Stage 1, and Stage 2 representations. The Stage 2 projections show the clearest visual alignment between the embedding clusters and the class colors, with substantially better separation than in the original representation. This qualitative structure indicates that Stage 2 produces embeddings that organize the variable-star classes more effectively.

\begin{figure*}[pos=!tp]
\centering
\includegraphics[width=\linewidth]{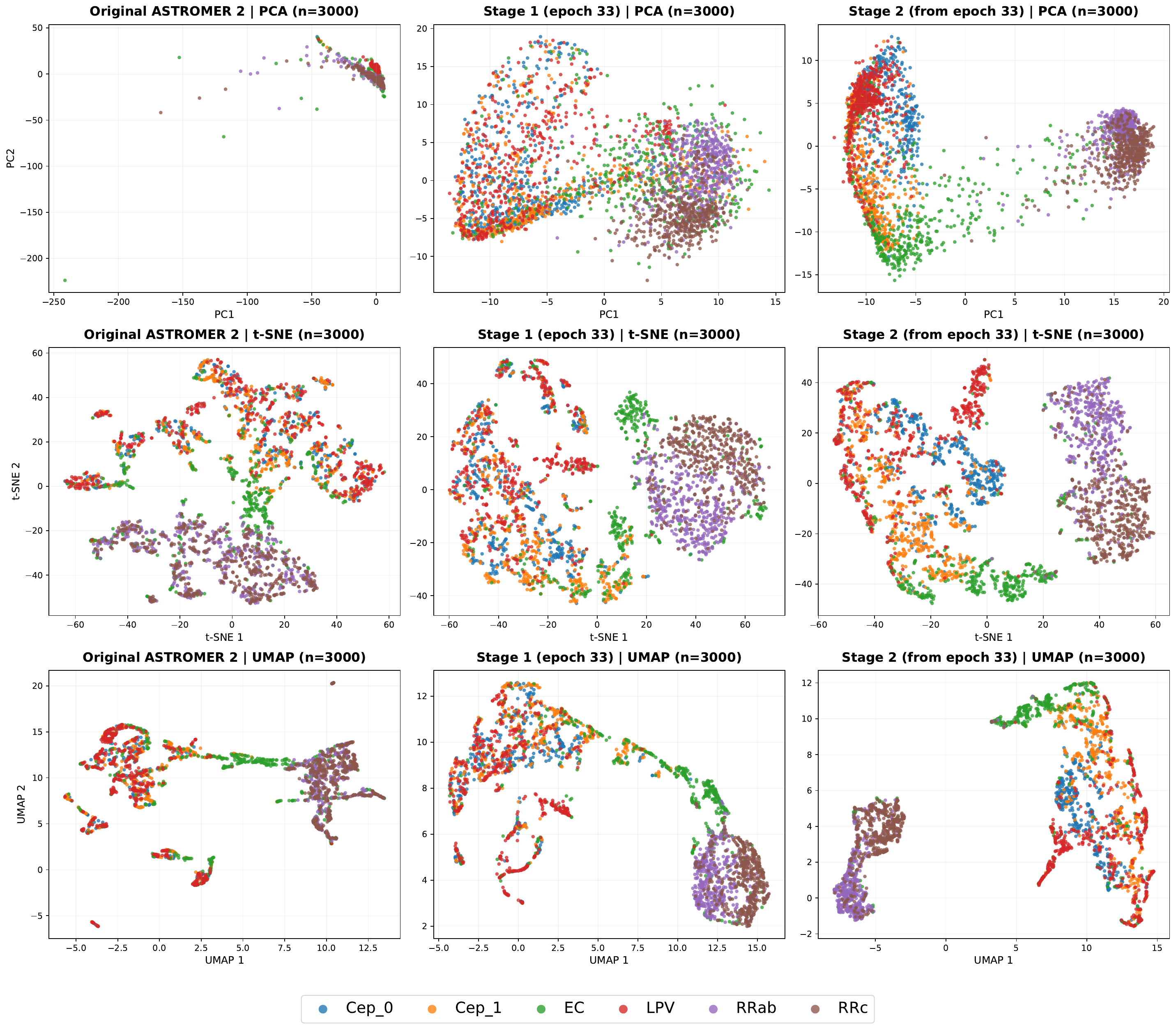}
\caption{Two-dimensional views of the 256-dimensional pooled-\(h\) representations from the original ASTROMER~2, Stage 1 epoch 33, and Stage 2 from epoch 33 encoders (columns), using PCA, t-SNE, and UMAP (rows). Every panel contains the same class-balanced sample of 3,000 objects, with 500 objects per class, drawn from the complete 20,894-object Alcock pool. This combined sample is used only for qualitative visualization and is not treated as held-out evaluation evidence. Projection coordinates have no direct physical interpretation and should not be compared numerically across panels.}
\label{fig:embedding_projection_grid}
\par\smallskip
\end{figure*}

We next quantify class-aligned organization directly in the original 256-dimensional pooled-\(h\) space. Unlike the visualization sample in Figure~\ref{fig:embedding_projection_grid}, all quantitative diagnostics are evaluated on the fixed 600-object locked holdout, containing 100 objects per class. For each encoder, a StandardScaler is fitted on the corresponding 3,000-object Alcock500 development pool and applied unchanged to the holdout embeddings. No downstream MLP is trained or used in this analysis.

We report three complementary diagnostics. First, k-nearest-neighbor (k-NN; k = 5) macro-F1 measures local label consistency by fitting a Euclidean k-NN classifier on the 3,000 labeled development embeddings and evaluating it on the locked holdout. Second, the inter/intra ratio divides the mean Euclidean distance between objects from different classes by the mean distance between objects from the same class; values above one indicate that between-class distances are larger on average. Third, the silhouette coefficient compares each object's average distance to its own class with its distance to the nearest other class~\cite{rousseeuw1987silhouettes}. The known classes define the reference groups for all three diagnostics, so these are label-referenced measures of embedding geometry rather than a fully unsupervised clustering experiment.

\begin{table*}[pos=!tp]
\centering
\caption{Direct embedding-space diagnostics evaluated on the fixed 600-object locked holdout. The k-NN classifier uses the 3,000-object development pool as its labeled reference set. Higher values are better for all three metrics.}
\label{tab:embedding_quality}
\begin{tabular}{lccc}
\toprule
Encoder & k-NN macro-F1 & Inter/intra ratio & Silhouette coefficient \\
\midrule
Original ASTROMER~2 & 0.4926 & 1.5267 & -0.1008 \\
Stage 1, epoch 33 & 0.7133 & 1.2038 & 0.0101 \\
Stage 2 from epoch 33 & \textbf{0.7462} & \textbf{1.7577} & \textbf{0.0926} \\
\bottomrule
\end{tabular}
\end{table*}

Table~\ref{tab:embedding_quality} shows that Stage 1 substantially improves local class consistency: k-NN macro-F1 increases from \(0.4926\) for the original ASTROMER~2 representation to \(0.7133\), while the silhouette coefficient changes from \(-0.1008\) to \(0.0101\). The inter/intra ratio decreases from \(1.5267\) to \(1.2038\), however, indicating that Stage 1 does not improve every global separation statistic. This distinction suggests that unsupervised contrastive learning produces more label-consistent local neighborhoods without yet organizing all classes into uniformly compact and separated global groups.

Stage 2 produces the strongest overall class-aware geometry. Relative to the matched Stage 1 encoder, its k-NN macro-F1 increases from \(0.7133\) to \(0.7462\), its inter/intra ratio increases from \(1.2038\) to \(1.7577\), and its silhouette coefficient increases from \(0.0101\) to \(0.0926\). The absolute silhouette value remains modest, indicating that some class overlap persists, but the agreement among the neighborhood, distance-ratio, silhouette, and projection diagnostics supports the conclusion that supervised contrastive adaptation produces the most class-discriminative pooled representation.


\section{Discussion}
\label{sec:discussion}

The classification and embedding-space results show that the two contrastive stages play complementary roles. Stage 1 makes the pooled ASTROMER~2 representation more useful for classification, while Stage 2 strengthens class organization and achieves the highest reported holdout macro-F1 across the three downstream label budgets. Together, these findings support adapting the representation-learning objective to improve the utility of a fixed encoder architecture.

\subsection{Contrastive Learning and Pooled Representations}
\label{subsec:discussion_why_contrastive}
\label{subsec:discussion_stage1_stage2}

ASTROMER~2 learns token representations by reconstructing masked observations from the surrounding light curve. This objective does not explicitly encourage the average of those representations to distinguish between variable-star classes. Stage 1 trains the encoder to bring representations of augmented views of the same object closer together and separate those of different objects. The gains in Table~\ref{tab:main_results} show that Stage 1 makes class information more accessible through this simple pooling operation.

Stage 2 then uses class labels to expand the positive set from another view of the same object to views of other objects in the same class. Together with the auxiliary cross-entropy objective on \(h\), this training encourages a representation that supports class discrimination. Stage 2 improves on its Stage 1 epoch-33 starting point at every downstream label budget, with the largest gain in Alcock20. This pattern is consistent with supervised contrastive adaptation making the frozen representation easier to use when the downstream classifier has limited labeled data.

\subsection{Embedding Geometry and Minority Classes}
\label{subsec:discussion_geometry_minority}

The direct embedding diagnostics provide evidence of improved class organization beyond the downstream MLP results. Stage 1 improves nearest-neighbor classification even though its inter/intra distance ratio decreases (Table~\ref{tab:embedding_quality}). This combination suggests that useful local class neighborhoods can emerge without a corresponding improvement in global separation. Stage 2 achieves the highest \(k\)-NN macro-F1, inter/intra ratio, and silhouette coefficient (Table~\ref{tab:embedding_quality}), linking its classification gains to stronger class organization in the pooled representation itself. The modest silhouette coefficient of 0.0926 indicates that the classes still overlap. The qualitative projections in Figure~\ref{fig:embedding_projection_grid} complement these full-dimensional diagnostics.

The class-level results show that Stage 2 also benefits the Cepheid classes, which are underrepresented in the Stage 2 training data. In Alcock500, both Cepheid classes improve after Stage 2, with the larger gain for the less abundant \texttt{Cep\_1} class. These gains show that the aggregate improvement includes better discrimination of minority classes despite the imbalance in the Stage 2 training data. Because the downstream classifier uses a balanced training sample, this finding concerns the usefulness of representations learned from imbalanced Stage 2 training data. It is particularly relevant to astronomical applications in which less abundant variability classes remain scientifically valuable.

\subsection{Comparison with the Official ASTROMER 2 Pipeline}
\label{subsec:discussion_pooled_vs_official}
\label{subsec:discussion_fairness}

The original ASTROMER~2 encoder's weak performance under simple averaging does not, by itself, show that its representations are poor. The official pipeline achieves much better classification performance with additional finetuning and a readout that combines information from multiple Transformer layers. Our results show that contrastive training makes the encoder's representations more useful for classification when token representations are simply averaged.

\subsection{Future Work}
\label{subsec:discussion_limitations}
\label{subsec:discussion_future_work}

This study focuses on making the pooled representation of a pretrained light-curve encoder more useful for classification. By retaining the ASTROMER~2 architecture and comparing frozen encoders through a common pooling rule and downstream classifier, we connect the observed gains to changes in the learned representation. The Stage~2 results establish the effectiveness of supervised contrastive learning combined with auxiliary cross-entropy. Separating these two training signals would address a complementary question: how much each contributes to class discrimination and embedding geometry.

Transfer to new variability classes would test how broadly the learned representation can be reused. In the present study, Stage~2 uses a separate labeled adaptation pool containing the same six classes as the downstream task; the reported label budgets specify the examples available to each downstream classifier. Evaluating classes excluded from Stage~2 would require a different partition of the class vocabulary, extending the current study from improving discrimination among known classes to transferring representations beyond the adaptation labels.

Richer readouts could reveal additional class information distributed across the encoder. Applying the same multilayer readout to the original, Stage~1, and Stage~2 encoders would test whether combining information from different depths further improves classification. Partial encoder finetuning would examine how these representations respond to additional task-specific training. These experiments would explore the interaction between representation learning and downstream model design, extending our present comparison of fixed, final-layer pooled representations.

Reliable uncertainty estimates could turn improved classification into better decisions about follow-up observations. A Gaussian process classifier applied to the pooled representations offers one route toward this goal~\cite{rasmussen2006gaussian}. Its predictive uncertainties would need to be assessed for calibration and for their ability to identify ambiguous light curves and prioritize follow-up targets. This direction introduces a decision-making objective beyond the classification performance and embedding geometry evaluated here.

Cross-survey evaluation would test whether the learned class structure persists under different observing conditions. The present experiments use single-band MACHO/Alcock light curves and a fixed augmentation family. Applying the method to surveys such as ZTF and ATLAS would introduce changes in cadence, noise, passbands, and class composition~\cite{bellm2019ztf,tonry2018atlas}. Evaluating transfer under these changes would also help determine whether the augmentations should be tailored to the target survey. Class-level performance and embedding-neighborhood analyses would reveal which benefits, including those for underrepresented classes, extend beyond Alcock.

\section{Conclusion}
\label{sec:conclusion}

Contrastive Astromer demonstrates that the pooled representation of a pretrained light-curve encoder can be made substantially more useful for variable-star classification while preserving its architecture. Starting from ASTROMER~2, unsupervised contrastive learning on MACHO encourages consistency across augmented views, and supervised contrastive adaptation with auxiliary cross-entropy on Alcock makes class distinctions more accessible in the resulting representation. The two stages connect learning from unlabeled light curves with the construction of a class-aware encoder that can be reused across downstream label budgets.

On Alcock, the adapted encoder achieves holdout macro-F1 scores of 0.7078, 0.7704, and 0.7943 with 20, 100, and 500 labeled examples per class for downstream classifier training after the separate labeled adaptation stage. These are the highest reported scores among the evaluated pipelines. Under the common frozen pooled-\(h\) interface, Stage 1 substantially improves on the original encoder, and Stage 2 adds a further 0.0514--0.1022 macro-F1 over its matched Stage 1 initialization. The largest Stage 2 gain occurs at the smallest downstream label budget.

The benefits extend to the organization of the representation itself. Direct embedding diagnostics show stronger local class consistency and class separation after Stage 2, while the Cepheid results demonstrate improved discrimination of classes underrepresented in the adaptation data. Classification, embedding geometry, and class-level performance therefore provide complementary evidence that contrastive adaptation makes the pooled representation more informative about variability class. Contrastive Astromer provides a concrete path from reconstruction-pretrained light-curve encoders to compact representations that expose class structure for both prediction and direct analysis.

\section*{Funding}
This research received no specific grant from public, commercial, or not-for-profit funding agencies.

\bibliographystyle{cas-model2-names}
\bibliography{references}

@article{donoso2023astromer,
  author = {Donoso-Oliva, C. and Becker, I. and Protopapas, P. and Cabrera-Vives, G. and Vishnu, M. and Vardhan, H.},
  title = {{ASTROMER}: A Transformer-Based Embedding for the Representation of Light Curves},
  journal = {Astronomy \& Astrophysics},
  volume = {670},
  pages = {A54},
  year = {2023},
  doi = {10.1051/0004-6361/202243928},
  eprint = {2205.01677},
  archivePrefix = {arXiv},
  primaryClass = {astro-ph.IM}
}

@article{donoso2026astromer2,
  author = {Donoso-Oliva, Cristobal and Becker, Ignacio and Protopapas, Pavlos and Cabrera-Vives, Guillermo and C{\'a}diz-Leyton, Martina and Moreno-Cartagena, Daniel},
  title = {Generalizing Across Astronomical Surveys: Few-Shot Light Curve Classification with {Astromer 2}},
  journal = {Astronomy \& Astrophysics},
  volume = {707},
  pages = {A170},
  year = {2026},
  doi = {10.1051/0004-6361/202554026},
  eprint = {2502.02717},
  archivePrefix = {arXiv},
  primaryClass = {astro-ph.IM}
}

@misc{astromer2zenodo,
  author = {Protopapas, Pavlos and Cabrera-Vives, Guillermo and Becker, Ignacio and Moreno Cartagena, Daniel and C{\'a}diz-Leyton, Martina},
  title = {Supplementary Material for: Generalizing Across Astronomical Surveys: Few-Shot Light Curve Classification with {Astromer 2}},
  howpublished = {Zenodo dataset},
  year = {2026},
  version = {2.0.0},
  doi = {10.5281/zenodo.18207945},
  url = {https://zenodo.org/records/18207945}
}

@article{bommasani2021opportunities,
  author = {Bommasani, Rishi and Hudson, Drew A. and Adeli, Ehsan and Altman, Russ and Arora, Simran and von Arx, Sydney and Bernstein, Michael S. and Bohg, Jeannette and Bosselut, Antoine and Brunskill, Emma and others},
  title = {On the Opportunities and Risks of Foundation Models},
  journal = {arXiv preprint arXiv:2108.07258},
  year = {2021},
  url = {https://arxiv.org/abs/2108.07258}
}

@article{oord2018representation,
  author = {van den Oord, Aaron and Li, Yazhe and Vinyals, Oriol},
  title = {Representation Learning with Contrastive Predictive Coding},
  journal = {arXiv preprint arXiv:1807.03748},
  year = {2018},
  url = {https://arxiv.org/abs/1807.03748}
}

@inproceedings{chen2020simple,
  author = {Chen, Ting and Kornblith, Simon and Norouzi, Mohammad and Hinton, Geoffrey},
  title = {A Simple Framework for Contrastive Learning of Visual Representations},
  booktitle = {Proceedings of the 37th International Conference on Machine Learning},
  series = {Proceedings of Machine Learning Research},
  volume = {119},
  pages = {1597--1607},
  year = {2020},
  publisher = {PMLR},
  url = {https://proceedings.mlr.press/v119/chen20j.html}
}

@inproceedings{khosla2020supervised,
  author = {Khosla, Prannay and Teterwak, Piotr and Wang, Chen and Sarna, Aaron and Tian, Yonglong and Isola, Phillip and Maschinot, Aaron and Liu, Ce and Krishnan, Dilip},
  title = {Supervised Contrastive Learning},
  booktitle = {Advances in Neural Information Processing Systems},
  volume = {33},
  pages = {18661--18673},
  year = {2020},
  url = {https://arxiv.org/abs/2004.11362}
}

@article{rousseeuw1987silhouettes,
  author = {Rousseeuw, Peter J.},
  title = {Silhouettes: A Graphical Aid to the Interpretation and Validation of Cluster Analysis},
  journal = {Journal of Computational and Applied Mathematics},
  volume = {20},
  pages = {53--65},
  year = {1987},
  doi = {10.1016/0377-0427(87)90125-7}
}

@article{vandermaaten2008visualizing,
  author = {van der Maaten, Laurens and Hinton, Geoffrey},
  title = {Visualizing Data Using {t-SNE}},
  journal = {Journal of Machine Learning Research},
  volume = {9},
  number = {86},
  pages = {2579--2605},
  year = {2008},
  url = {https://www.jmlr.org/papers/v9/vandermaaten08a.html}
}

@article{mcinnes2018umap,
  author = {McInnes, Leland and Healy, John and Saul, Nathaniel and Gro{\ss}berger, Lukas},
  title = {{UMAP}: Uniform Manifold Approximation and Projection},
  journal = {Journal of Open Source Software},
  volume = {3},
  number = {29},
  pages = {861},
  year = {2018},
  doi = {10.21105/joss.00861}
}

@inproceedings{xie2016deep,
  author = {Xie, Junyuan and Girshick, Ross and Farhadi, Ali},
  title = {Unsupervised Deep Embedding for Clustering Analysis},
  booktitle = {Proceedings of the 33rd International Conference on Machine Learning},
  series = {Proceedings of Machine Learning Research},
  volume = {48},
  pages = {478--487},
  year = {2016},
  publisher = {PMLR},
  url = {https://proceedings.mlr.press/v48/xieb16.html}
}

@inproceedings{mitrovic2020less,
  author = {Mitrovic, Jovana and McWilliams, Brian and Rey, Melanie},
  title = {Less Can Be More in Contrastive Learning},
  booktitle = {Proceedings on ``I Can't Believe It's Not Better!'' at NeurIPS Workshops},
  series = {Proceedings of Machine Learning Research},
  volume = {137},
  pages = {70--75},
  year = {2020},
  publisher = {PMLR},
  url = {https://proceedings.mlr.press/v137/mitrovic20a.html}
}

@inproceedings{lanusse2023astroclip,
  author = {Lanusse, Francois and Parker, Liam Holden and Golkar, Siavash and Bietti, Alberto and Cranmer, Miles and Eickenberg, Michael and Krawezik, Geraud and McCabe, Michael and Ohana, Ruben and Pettee, Mariel and others},
  title = {{AstroCLIP}: Cross-Modal Pre-Training for Astronomical Foundation Models},
  booktitle = {NeurIPS 2023 AI for Science Workshop},
  year = {2023},
  eprint = {2310.03024},
  archivePrefix = {arXiv},
  primaryClass = {astro-ph.IM},
  url = {https://openreview.net/forum?id=Zt35qMfDTp}
}

@article{parker2024astroclip,
  author = {Parker, Liam and Lanusse, Francois and Golkar, Siavash and Sarra, Leopoldo and Cranmer, Miles and Bietti, Alberto and Eickenberg, Michael and Krawezik, Geraud and McCabe, Michael and Morel, Rudy and Ohana, Ruben and Pettee, Mariel and Regaldo-Saint Blancard, Bruno and Cho, Kyunghyun and Ho, Shirley and Polymathic AI Collaboration},
  title = {{AstroCLIP}: A Cross-Modal Foundation Model for Galaxies},
  journal = {Monthly Notices of the Royal Astronomical Society},
  volume = {531},
  number = {4},
  pages = {4990--5011},
  year = {2024},
  doi = {10.1093/mnras/stae1450},
  eprint = {2310.03024},
  archivePrefix = {arXiv},
  primaryClass = {astro-ph.IM}
}

@article{zhang2024maven,
  author = {Zhang, Gemma and Helfer, Thomas and Gagliano, Alexander T. and Mishra-Sharma, Siddharth and Villar, V. Ashley},
  title = {{Maven}: A Multimodal Foundation Model for Supernova Science},
  journal = {Machine Learning: Science and Technology},
  volume = {5},
  number = {4},
  pages = {045069},
  year = {2024},
  doi = {10.1088/2632-2153/ad990d},
  eprint = {2408.16829},
  archivePrefix = {arXiv},
  primaryClass = {astro-ph.IM}
}

@article{ding2026starclr,
  author = {Ding, Junyao and Chen, Xiaodian and Gao, Xinyi and Tang, Xiaoyu and Wang, Shu and Huang, Yang and Qi, Xinyu and Xue, Guirong and Luo, Ali and Liu, Jifeng},
  title = {{StarCLR}: Contrastive Learning Representation for Astronomical Light Curves},
  journal = {The Astrophysical Journal},
  volume = {1003},
  pages = {141},
  year = {2026},
  doi = {10.3847/1538-4357/ae64ef},
  eprint = {2604.24516},
  archivePrefix = {arXiv},
  primaryClass = {astro-ph.IM}
}

@article{rizhko2025astrom3,
  author = {Rizhko, Mariia and Bloom, Joshua S.},
  title = {{AstroM}$^{3}$: A Self-Supervised Multimodal Model for Astronomy},
  journal = {The Astronomical Journal},
  volume = {170},
  pages = {28},
  year = {2025},
  doi = {10.3847/1538-3881/adcbad},
  eprint = {2411.08842},
  archivePrefix = {arXiv},
  primaryClass = {astro-ph.IM}
}

@book{rasmussen2006gaussian,
  author = {Rasmussen, Carl Edward and Williams, Christopher K. I.},
  title = {Gaussian Processes for Machine Learning},
  series = {Adaptive Computation and Machine Learning},
  publisher = {MIT Press},
  address = {Cambridge, MA},
  year = {2006},
  isbn = {026218253X},
  url = {https://gaussianprocess.org/gpml/}
}

@misc{abadi2015tensorflow,
  author = {Abadi, Mart{\'i}n and Agarwal, Ashish and Barham, Paul and Brevdo, Eugene and Chen, Zhifeng and Citro, Craig and Corrado, Greg S. and Davis, Andy and Dean, Jeffrey and Devin, Matthieu and Ghemawat, Sanjay and Goodfellow, Ian and Harp, Andrew and Irving, Geoffrey and Isard, Michael and Jia, Yangqing and Jozefowicz, Rafal and Kaiser, Lukasz and Kudlur, Manjunath and Levenberg, Josh and Man{\'e}, Dan and Monga, Rajat and Moore, Sherry and Murray, Derek and Olah, Chris and Schuster, Mike and Shlens, Jonathon and Steiner, Benoit and Sutskever, Ilya and Talwar, Kunal and Tucker, Paul and Vanhoucke, Vincent and Vasudevan, Vijay and Vi{\'e}gas, Fernanda and Vinyals, Oriol and Warden, Pete and Wattenberg, Martin and Wicke, Martin and Yu, Yuan and Zheng, Xiaoqiang},
  title = {{TensorFlow}: Large-Scale Machine Learning on Heterogeneous Systems},
  year = {2015},
  howpublished = {Software available from tensorflow.org},
  url = {https://www.tensorflow.org/}
}

@misc{chollet2015keras,
  author = {Chollet, Fran{\c{c}}ois and others},
  title = {Keras},
  year = {2015},
  howpublished = {\url{https://keras.io}}
}

@article{bellm2019ztf,
  author = {Bellm, Eric C. and others},
  title = {The {Zwicky Transient Facility}: System Overview, Performance, and First Results},
  journal = {Publications of the Astronomical Society of the Pacific},
  volume = {131},
  number = {995},
  pages = {018002},
  year = {2019},
  doi = {10.1088/1538-3873/aaecbe},
  eprint = {1902.01932},
  archivePrefix = {arXiv},
  primaryClass = {astro-ph.IM}
}

@article{tonry2018atlas,
  author = {Tonry, J. L. and Denneau, L. and Heinze, A. N. and Stalder, B. and Smith, K. W. and Smartt, S. J. and Stubbs, C. W. and Weiland, H. J. and Rest, A.},
  title = {{ATLAS}: A High-Cadence All-Sky Survey System},
  journal = {Publications of the Astronomical Society of the Pacific},
  volume = {130},
  number = {988},
  pages = {064505},
  year = {2018},
  doi = {10.1088/1538-3873/aabadf},
  eprint = {1802.00879},
  archivePrefix = {arXiv},
  primaryClass = {astro-ph.IM}
}

@article{eyer2008variable,
  author = {Eyer, Laurent and Mowlavi, Nami},
  title = {Variable Stars Across the Observational {HR} Diagram},
  journal = {Journal of Physics: Conference Series},
  volume = {118},
  number = {1},
  pages = {012010},
  year = {2008},
  doi = {10.1088/1742-6596/118/1/012010},
  eprint = {0712.3797},
  archivePrefix = {arXiv},
  primaryClass = {astro-ph}
}

@article{debosscher2007automated,
  author = {Debosscher, J. and Sarro, L. M. and Aerts, C. and Cuypers, J. and Vandenbussche, B. and Garrido, R. and Solano, E.},
  title = {Automated Supervised Classification of Variable Stars. {I}. {Methodology}},
  journal = {Astronomy \& Astrophysics},
  volume = {475},
  number = {3},
  pages = {1159--1183},
  year = {2007},
  doi = {10.1051/0004-6361:20077638}
}

@misc{cadizleyton2025astromoe,
  author = {C{\'a}diz-Leyton, Martina and Cabrera-Vives, Guillermo and Protopapas, Pavlos and Moreno-Cartagena, Daniel and Becker, Ignacio},
  title = {{Astro-MoE}: Mixture of Experts for Multiband Astronomical Time Series},
  howpublished = {arXiv preprint arXiv:2507.12611},
  year = {2025},
  doi = {10.48550/arXiv.2507.12611},
  eprint = {2507.12611},
  archivePrefix = {arXiv},
  primaryClass = {astro-ph.IM},
  url = {https://arxiv.org/abs/2507.12611}
}

\appendix
\setcounter{table}{0}
\renewcommand{\thetable}{\thesection.\arabic{table}}

\section{Implementation and Data-Audit Details}

\subsection{Mask Convention}

The contrastive wrapper uses a visible-observation convention: \(q_t=1\) denotes an observed token and \(q_t=0\) denotes padding or an augmentation-dropped token.  The official ASTROMER 2 attention implementation uses the opposite blocking convention internally.  Before calling the official encoder, the wrapper therefore passes \(1-q_t\) as the attention mask.  This conversion keeps the visible-observation convention used for contrastive pooling consistent with the blocking convention expected by the official attention layer.

\subsection{Data Package Provenance and Split Audits}

The object-ID audit verified that the three Alcock partitions in Table~\ref{tab:alcock_split_sizes} are mutually exclusive. A separate MACHO-to-Alcock audit removed 453 overlapping objects from the Stage 1 contrastive dataset before training.

All expected downstream runs and pooled-\(h\) and official-pipeline controls completed successfully.

\subsection{Embedding-Projection Configuration}
\label{app:projection_settings}

Figure~\ref{fig:embedding_projection_grid} uses a class-balanced sample drawn from the complete Alcock collection under the revised disjoint protocol. The source pool combines the three partitions in Table~\ref{tab:alcock_split_sizes}. Using random seed 42, we sample 500 objects without replacement from each of the six classes, producing a 3,000-object visualization set. Per-class reservoir sampling is followed by a deterministic global shuffle. The sampled records and their ordering are identical for the original ASTROMER~2, Stage 1 epoch 33, and Stage 2 from epoch 33 encoders.

For each encoder independently, the 256 embedding coordinates are standardized to zero mean and unit variance across the same 3,000 sampled objects before dimensionality reduction. PCA, t-SNE, and UMAP are then fitted independently to each encoder's standardized embeddings. Class labels are not supplied to any projection method and are used only after projection to color the points. All stochastic procedures use random seed 42. Table~\ref{tab:projection_settings} reports the projection configuration.

\begin{table}[pos=!htbp]
\centering
\caption{Sampling, preprocessing, and dimensionality-reduction settings used to generate Figure~\ref{fig:embedding_projection_grid}. Parameters not listed use the defaults of the recorded software versions.}
\label{tab:projection_settings}
\begin{tabular}{@{}>{\raggedright\arraybackslash}p{0.25\linewidth} >{\raggedright\arraybackslash}p{0.65\linewidth}@{}}
\toprule
Method & Configuration \\
\midrule
Sampling & 500 objects per class sampled without replacement from 20,894 unique Alcock objects; 3,000 objects total; random seed 42. \\
Preprocessing & StandardScaler fitted independently to each encoder's 3,000 sampled embeddings. \\
PCA & Two components; random seed 42. \\
t-SNE & Two components; perplexity 30; PCA initialization; automatic learning rate; random seed 42. \\
UMAP & Two components; 15 nearest neighbors; minimum distance 0.1; Euclidean distance; random seed 42. \\
\bottomrule
\end{tabular}
\end{table}

The projections were generated using Python 3.10.20, NumPy 1.26.4, scikit-learn 1.7.2, Matplotlib 3.10.9, and UMAP-learn 0.5.12. Because each projection is fitted independently for each encoder, the coordinate values, orientations, and scales should not be compared numerically across panels. The projections are used only to compare visible class organization within each representation space.

\end{document}